\documentclass[final]{article}

\usepackage[top=2.3cm, bottom=2cm, left=3cm, right=3cm, a4paper]{geometry}

\usepackage[utf8]{inputenc}
\usepackage[T1]{fontenc}
\usepackage[english]{babel}

\usepackage{amsmath,amsfonts,amssymb,amsthm,mathrsfs,mathabx}

\theoremstyle{plain}
\newtheorem{theorem}{Theorem}[section]

\theoremstyle{definition}

\usepackage{cleveref}

\crefname{lemma}{Lemma}{Lemmas}
\crefname{prop}{Proposition}{Propositions}
\crefname{cor}{Corollary}{Corollaries}
\crefname{thm}{Theorem}{Theorems}
\crefname{defi}{Definition}{Definitions}
\crefname{rem}{Remark}{Remarks}
\crefname{ex}{Example}{Examples}
\crefname{obs}{Observation}{Observations}

\newcommand{\N}{\mathbb{N}}

\newcommand{\eps}{\varepsilon}

\newcommand{\floor}[1]{\left \lfloor #1 \right \rfloor}
\newcommand{\set}[1]{\left \{ #1 \right \}}

\usepackage[pdftex]{graphicx}
\usepackage[usenames,dvipsnames,table]{xcolor}
\definecolor{shade}{RGB}{235,235,235}
\usepackage{hyperref}
\usepackage{cleveref}

\usepackage{todonotes}

\title{Additive spanners: A simple construction}
\author{Mathias B\ae k Tejs Knudsen\thanks{Research partly supported by
Thorup's Advanced Grant from the Danish Council for Independent
Research under the Sapere Aude research carrier programme and by the
FNU project AlgoDisc - Discrete Mathematics, Algorithms, and Data
Structrues}
\\
University of Copenhagen
}

\begin{document}

\maketitle

\begin{abstract}
We consider additive spanners of unweighted undirected graphs.
Let $G$ be a graph and $H$ a subgraph of $G$.
The most naïve way to construct an additive $k$-spanner of $G$ is the following:
As long as $H$ is not an additive $k$-spanner repeat:
Find a pair $(u,v) \in H$ that violates the spanner-condition and a shortest path from $u$ to $v$ in $G$. Add the edges of this path to $H$.

We show that, with a very simple initial graph $H$, this naïve method gives additive $6$- and $2$-spanners of sizes matching the best known upper bounds.
For additive $2$-spanners we start with $H=\emptyset$ and end with $O(n^{3/2})$ edges in the spanner. For additive $6$-spanners we start with $H$ containing $\lfloor n^{1/3} \rfloor$ arbitrary edges incident to each node and end with a spanner of size $O(n^{4/3})$.
\end{abstract}
\section{Introduction}
Additive spanners are subgraphs that preserve the distances in the graph up to an additive positive constant. Given an unweighted undirected graph $G$, a subgraph $H$ is an additive $k$-spanner if for every pair of nodes $u,v$ it is true that
\[
d_G(u,v) \le d_H(u,v) \le d_G(u,v)+k
\]
In this paper we only consider purely additive spanners, which are $k$-spanners where $k = O(1)$. Throughout this paper every graph will be unweighted and undirected.


Many people have considered a variant of this problem, namely multiplicative spanners and even mixes between additive and multiplicative spanners \cite{pettie07spanners,elkin01spanners,thorup06spanners}.
The problem of finding a $k$-spanner of smallest size has received a lot of attention. Most notably, given a graph with $n$ nodes Dor et al. \cite{dor00apsp} prove that it has a $2$-spanner of size $O(n^{3/2})$, Baswana et al. \cite{baswana05spanners} prove that it has a $6$-spanner of size $O(n^{4/3})$, and Chechik \cite{chechik13spanners} proves that it has a $4$-spanner of size $O(n^{7/5}\log^{1/5}n)$. Woodrufff \cite{woodruff06spanners} shows that for every constant $k$ there exist graphs with $n$ nodes such that every $(2k-1)$-spanner must have at least $\Omega(n^{1+1/k})$ edges. This implies that the construction of $2$-spanners are optimal.
Whether there exists an algorithm for constructing $O(1)$-spanners with $O(n^{1+\eps})$ edges for some $\eps < 1/3$ is unknown and is an important open problem.

Let $G$ be a graph and $H$ a subgraph of $G$. Consider the following algorithm: As long as there exists a pair of nodes $u,v$ such that $d_H(u,v) > d_G(u,v) + k$, find a shortest path from $u$ to $v$ in $G$ and add the edges on the path to $H$. This process will be referred to as \textbf{$k$-spanner-completion}. After $k$-spanner-completion, $H$ will be a $k$-spanner of $G$.
Thus, given a graph $G$, a general way to construct a $k$-spanner for $G$ is the following: Firstly, find a simple subgraph of $G$. Secondly use $k$-spanner-completion on this subgraph. The main contribution of this paper is:

\begin{theorem}
\label{6-spanner-thm}
Let $G$ be a graph with $n$ nodes and $H$ the subgraph containing all nodes but no edges of $G$. For each node add $\floor{n^{1/3}}$ edges adjacent to that node to $H$ (or, if the degree is less, add all edges incident to that node). After $6$-spanner-completion $H$ will have at most $O(n^{4/3})$ edges.
\end{theorem}

It is well-known that a graph with $n$ nodes has a $6$-spanner of size $O(n^{4/3})$ \cite{baswana05spanners}. The techniques employed in our proof of correctness are similar to those in \cite{baswana05spanners}. The creation of the initial graph $H$ corresponds to the clustering in \cite{baswana05spanners} and the $6$-spanner-completion corresponds to their path-buying algorithm. For completeness we show that the same method gives a $2$-spanner of size $O(n^{3/2})$. This fact is already known due to \cite{dor00apsp} and is matched by a lower bound from \cite{woodruff06spanners}.

\begin{theorem}
\label{2-spanner-thm}
Let $G$ be a graph with $n$ nodes and $H$ the subgraph where all edges are removed. Upon $2$-spanner-completion $H$ has at most $O(n^{3/2})$ edges.
\end{theorem}
\section{Creating a 6-spanner}

The algorithm for creating a $6$-spanner was described in the abstract and the introduction.

For a given graph $G$, a $6$-spanner of $G$ can be created by strating with some subgraph $H$ of $G$ and applying $6$-spanner-completion to $H$. \Cref{6-spanner-thm} states that for a suitable starting choice of $H$ we get a spanner of size $O(n^{4/3})$.
The purpose of this section is to show that the $6$-spanner created has no more than $O(n^{4/3})$ edges. This will imply that the construction (in terms of the size of the $6$-spanner) matches the best known upper bound \cite{baswana05spanners}.

\begin{proof}[of \Cref{6-spanner-thm}]
Inserting (at most) $\floor{n^{1/3}}$ edges per node will only add $n\floor{n^{1/3}} = O(n^{4/3})$ edges to $H$. Therefore it is only necessary to prove that $6$-spanner-completion adds no more than $O(n^{4/3})$ edges.

Let $v(H)$ and $c(H)$ be defined by:
\[
v(H) = 
\sum_{u,v \in V(G)} 
\max \set{0, d_G(u,v)-d_H(u,v)+5}
,
\quad 
c(H) = \# E(H)
\]
Say that a shortest path, $p$, from $u$ to $v$ is added to $H$, and let $H_0$ be the subgraph before the edges are added. Let the path consist of the nodes:
\[
u = w_0, w_1, \ldots, w_r = v, r \in \N
\]
Let $u' = w_i$ be the node $w_i$ with the smallest $i$ such that $\deg_{H_0}(w_i) \ge \floor{n^{1/3}}$. Likewise let $v' = w_j$ be the node $w_j$ the largest $j$ such that $\deg_{H_0}(w_j) \ge \floor{n^{1/3}}$. Remember that if $\deg_{H_0}(w_i) < \floor{n^{1/3}}$ then all the edges adjacent to $w_i$ are already in $H_0$. This implies that $d_{H_0}(u',v') > d_G(u',v')+6$ since $d_{H_0}(u,v) > d_G(u,v)+6$.

Say that $t$ new edges are added to $H$. Then there must be at least $t$ nodes on $p$ with degree $>n^{1/3}$. Since every node can be adjacent to no more than $3$ nodes on $p$ (since it is a shortest path) there must be $\Omega(n^{1/3}t)$ nodes adjacent to $p$ in $H$. Let $z$ and $w$ be neighbours to $u'$ and $v'$ in $H$ respectively and let $r$ be any node adjacent to $p$ in $H$. Let $s$ be a node on $p$ such that $r$ and $s$ are adjacent in $H$. See \Cref{6spanner-img} for an illustration.

\begin{figure}[htbp]
	\centering
	\includegraphics[width=.5\textwidth]{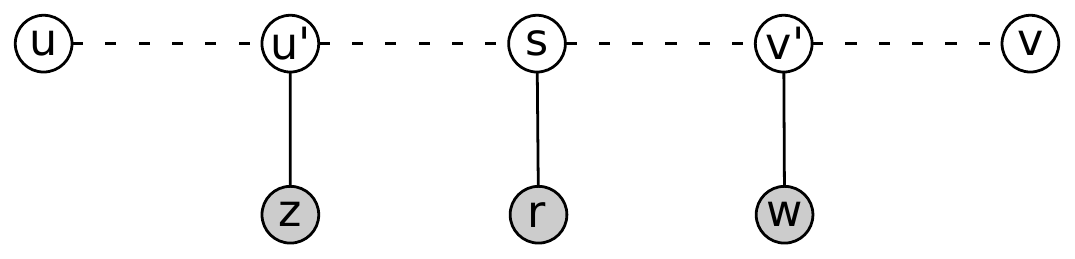}	
	\caption{The dashed line denotes the shortest path from $u$ to $v$. The solid lines denote edges.}
	\label{6spanner-img}
\end{figure}
By the triangle inequality we see that:
\begin{align*}
d_H(z,r) + d_H(r,w) \le d_G(u',v')+4
\end{align*}
But on the other hand:
\begin{align*}
d_{H_0}(z,r) + d_{H_0}(r,w) \ge 
d_{H_0}(z,w) \ge 
d_{H_0}(u',v') - 2 >
d_G(u',v') + 4
\end{align*}
Combining these two inequalities we obtain $d_{H_0}(z,r) > d_H(z,r)$ or $d_{H_0}(r,w) > d_H(r,w)$. And from the triangle inequality $d_G(z,r)+5 > d_H(z,r)$ and $d_G(r,w)+5 > d_H(r,w)$.
Since $u'$ and $v'$ have at least $n^{1/3}$ neighbours and there are $\Omega(n^{1/3}t)$ nodes in $H$ adjacent to $p$, the definition of $v(H)$ implies that:
\[
v(H) - v(H_0) \ge \Omega(t(n^{1/3})^2)
\]
And since $c(H) - c(H_0) = t$:
\[
\frac{v(H)-v(H_0)}{c(H)-c(H_0)} \ge 
\Omega(n^{2/3})
\]
Since $v(H) \le O(n^2)$ this implies that $c(H)$ increases with no more than \linebreak $O(n^2/n^{2/3}) = O(n^{4/3})$ in total when all shortest paths are inserted. Hence $c(H) = O(n^{4/3})$ when the $6$-spanner-completion is finished which yields the conclusion.
\end{proof}
\section{Creating a 2-spanner}
For completeness we show that $2$-spanner-completion gives spanners with $O(n^{3/2})$ edges.
This matches the upper bound from \cite{dor00apsp} and the lower bound from \cite{woodruff06spanners}.

\begin{proof}[of \Cref{2-spanner-thm}]
Let $G$ be a graph with $n$ nodes. Whenever $H$ is a spanner of $G$, define $v(H)$ and $c(H)$ as:
\[
v(H) = 
\sum_{u,v \in V(G)} 
\max \set{0, d_G(u,v) - d_H(u,v) + 3}
,
\quad
c(H) = \sum_{v \in V(G)} \left(\deg_H(v)\right)^2
\]
It is easy to see that $0 \le v(H) \le 3n^2$ and by Cauchy-Schwartz's inequality $\sqrt{c(H) \cdot n} \ge 2\# E(H)$. The goal will be to prove that when the algorithm terminates $c(H) = O(n^2)$, since this implies that $\# E(H) = O(n^{3/2})$. This is done by proving that in each step of the algorithm $c(H) - 12 v(H)$ will not increase. Since $v(H) = O(n^2)$ this means that $c(H) = O(n^2)$ which ends the proof. Therefore it is sufficient to check that $c(H) - 12v(H)$ never increases.

Consider a step where new edges are added to $H$ on a shortest path from $u$ to $v$ of length $t$. Let $H_0$ be the subgraph before the edges are added. Assume that $u,v$ violates the $2$-spanner condition in $H_0$, i.e.~$d_{H_0}(u,v) > d_G(u,v)+2$. Let the shortest path consist of the nodes:
\[
u = w_0, w_1, \ldots, w_{t-1}, w_t = v
\]
It is obvious that:
\begin{align*}
c(H) - c(H_0) \le \sum_{i=0}^t (\deg_H(w_i))^2-(\deg_H(w_i)-2)^2
\le 4\sum_{i=0}^t \deg_H(w_i)
\end{align*}
Every node cannot be adjacent to more than $3$ nodes on the shortest path, since otherwise it would not be a shortest path. Using this insight we can bound the number of nodes which in $H$ are adjacent to or on the shortest path from below by:
\[
\frac{1}{3} \sum_{i=0}^t \deg_H(w_i)
\]
Now let $z$ be a node in $H$ adjacent or on to the shortest path. Obviously:
\[
d_{H}(u,z) + d_{H}(z,v) \le 
d_G(u,v)+2
\]
Furthermore $d_{H_0}(u,z) + d_{H_0}(z,v) > d_G(u,v)+2$ since otherwise there would exist a path from $u$ to $v$ in $H_0$ of length $\le d_G(u,v)+2$. Hence:
\[
d_{H}(u,z) + d_{H}(z,v) < 
d_{H_0}(u,z) + d_{H_0}(z,v)
\]
Now let $z$ be a node on the shortest path which is adjacent to $w_i$ in $H$ (every node on the path will also be adjacent in $H$ to such a node). Then by the triangle inequality:
\begin{alignat*}{2}
d_H(u,z) & \le 
d_H(u,w_i) + d_H(w_i,z) & & =
d_G(u,w_i) + 1 \\ 
&\le
d_G(u,z) + d_G(z,w_i) + 1 & & =
d_G(u,z) + 2
\end{alignat*}
And likewise $d_H(z,v) \le d_G(z,v)+2$. Combining these two observations yields:
\[
\sum_{w \in V}
\max \set{0, d_G(z,w) - d_H(z,w) + 3}
<
\sum_{w \in V}
\max \set{0, d_G(z,w) - d_{H_0}(z,w) + 3}
\]
Since this holds for every node in $H$ adjacent to or on the shortest path this means that:
\[
v(H) - v(H_0) \ge 
\frac{1}{3} \sum_{i=0}^t \deg_H(w_i)
\]
Combining this with the bound on $c(H) - c(H_0)$ gives:
\[
(c(H) - 12 v(H)) - (c(H_0) - 12 v(H_0)) \le 0
\]
which finishes the proof.
\end{proof}

\bibliographystyle{plain}
\bibliography{refs}

\begin{thebibliography}{1}

\bibitem{baswana05spanners}
Surender Baswana, Telikepalli Kavitha, Kurt Mehlhorn, and Seth Pettie.
\newblock New constructions of {$(\alpha, \beta)$}-spanners and purely additive
  spanners.
\newblock In {\em Proc. 16th ACM/SIAM Symposium on Discrete Algorithms (SODA)},
  pages 672--681, 2005.

\bibitem{chechik13spanners}
Shiri Chechik.
\newblock New additive spanners.
\newblock In {\em Proc. 24th ACM/SIAM Symposium on Discrete Algorithms (SODA)},
  pages 498--512, 2013.

\bibitem{dor00apsp}
Dorit Dor, Shay Halperin, and Uri Zwick.
\newblock All-pairs almost shortest paths.
\newblock {\em SIAM Journal on Computing}, 29(5):1740--1759, 2000.
\newblock See also FOCS'96.

\bibitem{elkin01spanners}
Michael Elkin and David Peleg.
\newblock {$(1+\varepsilon, \beta)$}-spanner constructions for general graphs.
\newblock {\em SIAM Journal on Computing}, 33(3):608--631, 2004.
\newblock See also STOC'01.

\bibitem{pettie07spanners}
Seth Pettie.
\newblock Low distortion spanners.
\newblock In {\em Proc. 34th International Colloquium on Automata, Languages
  and Programming (ICALP)}, pages 78--89, 2007.

\bibitem{thorup06spanners}
Mikkel Thorup and Uri Zwick.
\newblock Spanners and emulators with sublinear distance errors.
\newblock In {\em Proc. 17th ACM/SIAM Symposium on Discrete Algorithms (SODA)},
  pages 802--809, 2006.

\bibitem{woodruff06spanners}
David~P. Woodruff.
\newblock Lower bounds for additive spanners, emulators, and more.
\newblock In {\em Proc. 47th IEEE Symposium on Foundations of Computer Science
  (FOCS)}, pages 389--398, 2006.

\end{thebibliography}

\end{document}